\documentclass[onecollarge,runningheads]{svjour2}
\smartqed  
\usepackage{graphicx}
\def\fdg{\hbox{$.\!\!^\circ$}}          
\newcommand{\etal}{et\,al.}

\newcommand{\degs}{\ifmmode ^{\circ}\else$^{\circ}$\fi}
\newbox\grsign \setbox\grsign=\hbox{$>$}
\newdimen\grdimen \grdimen=\ht\grsign
\newbox\laxbox \newbox\gaxbox
\setbox\gaxbox=\hbox{\raise.5ex\hbox{$>$}\llap
     {\lower.5ex\hbox{$\sim$}}}\ht1=\grdimen\dp1=0pt
\setbox\laxbox=\hbox{\raise.5ex\hbox{$<$}\llap
     {\lower.5ex\hbox{$\sim$}}}\ht2=\grdimen\dp2=0pt
\newcommand{\gax}{$\mathrel{\copy\gaxbox}$}
\newcommand{\lax}{$\mathrel{\copy\laxbox}$}
\journalname{Experimental Astronomy}
\begin{document}

\title{GRIPS\thanks{Excerpt of a proposal submitted in response to
the ESA Cosmic Vision Call in June 2007, invited for an audition on October 9
at ESA headquarters, but not selected for further
consideration in mid-October 2007.}
}
\subtitle{Gamma-Ray Burst Investigation via Polarimetry and Spectroscopy\\
www.grips-mission.eu\footnotemark}


\author{J. Greiner    \and
A. Iyudin  \and
G. Kanbach  \and
A. Zoglauer  \and
R. Diehl  \and
F. Ryde  \and
D. Hartmann  \and
A. v. Kienlin  \and
S. McBreen  \and
M. Ajello  \and
Z. Bagoly  \and
L.G. Balasz  \and
G. Barbiellini  \and
R. Bellazini  \and
L. Bezrukov  \and
D.V. Bisikalo  \and
G. Bisnovaty-Kogan  \and
S. Boggs  \and
A. Bykov  \and
A.M. Cherepashuk  \and
A. Chernenko  \and
W. Collmar \and
G. DiCocco  \and
W. Dr\"oge  \and
M. Gierlik  \and
L. Hanlon  \and
I. Horvath  \and
R. Hudec  \and
J. Kiener  \and
C. Labanti  \and
N. Langer  \and
S. Larsson  \and
G. Lichti  \and
V.M. Lipunov  \and
B.K. Lubsandorgiev  \and
A. Majczyna  \and
K. Mannheim  \and
R. Marcinkowski  \and
M. Marisaldi  \and
B. McBreen  \and
A. Meszaros  \and
E. Orlando  \and
M.I. Panasyuk  \and
M. Pearce  \and
E. Pian  \and
R.V. Poleschuk  \and
A. Pollo  \and
A. Pozanenko  \and
S. Savaglio  \and
B. Shustov  \and
A. Strong  \and
S. Svertilov  \and
V. Tatischeff  \and
J. Uvarov  \and
D.A. Varshalovich  \and
C.B. Wunderer  \and
G. Wrochna  \and
A.G. Zabrodskij  \and
L.M. Zeleny
\vspace{1.cm}
}

\authorrunning{Greiner et al.} 

\institute{Jochen Greiner \at
          Max-Planck-Institut f\"ur extraterrestrische Physik, 85740 Garching, Germany  \\
             Tel.: +49-89-30000-3847\\
             Fax: +49-89-30000-3606\\
             \email{jcg@mpe.mpg.de} 
}

\date{Received: 31 October 2007 / Accepted: date}

\maketitle

\begin{abstract}
The primary scientific goal of the GRIPS mission is to
revolutionize our understanding of the early universe using
$\gamma$-ray bursts.
We propose a new
generation gamma-ray observatory capable of unprecedented spectroscopy over a
wide range of $\gamma$-ray energies (200 keV--50 MeV) and of polarimetry
(200--1000 keV). Secondary goals achievable by this mission
include direct measurements of supernova interiors through $\gamma$-rays from
radioactive decays, nuclear astrophysics with massive stars and novae, and
studies of particle acceleration near compact stars, interstellar shocks, and
clusters of galaxies.

\end{abstract}

\footnotetext[1]{See this Web-site for the author's affiliations.}

\section{Introduction: Stepping beyond classical limits}

Gamma-ray bursts (GRB) are the most luminous sources in the sky, and
thus act as signposts throughout the Universe. The long-duration
sub-group is produced by the explosion of massive stars, while
short-duration GRBs likely originate
during the merging of compact objects.
Both types are intense neutrino sources, and being stellar sized
objects at cosmological scales, they connect different branches of
research and thus have a broad impact on present-day astrophysics.

Identifying objects at redshift \gax 6.5 has become
one of the main goals of modern observational cosmology, but turned
out to be difficult.
GRBs offer a promising opportunity
to  identify
high-$z$ objects, and moreover even allow us to
investigate
the host galaxies at these redshifts. GRBs are a factor
10$^{5-7}$
brighter than quasars during the first hour after explosion, and a
favourable relativistic
k-correction implies that they do not get fainter beyond $z$$\sim$3.
Yet, present and near-future ground- and space-based sensitivity limits
the measurement of redshifts at $z$$\sim$13 (as $H$-band drop-outs),
because GRB afterglows above 2.5 $\mu$m are
too faint by many magnitudes 
for 8--10\,m telescopes. 
Thus, a completely different
strategy is needed to step beyond redshift 13  to measure
when the first stars formed.

Fortunately, nuclear physics offers such a new strategy.
Similar to X-ray and optical absorption lines due to transitions
between electronic levels, resonant absorption processes in the nuclei
exist which leave narrow absorption lines in the $\gamma$-ray range.
The most prominent and astrophysically relevant are the nuclear excitation and 
Pygmy resonances (element-specific narrow lines between 5--9 MeV),
the Giant Dipole resonance  (GDR; proton versus neutron fluid oscillations;
$\sim$ 25 MeV; two nucleons and more)
and the Delta-resonance (individual-nucleon excitations,
325 MeV; all nucleons, including H!).
Such resonant absorption only depends on the presence of the
nucleonic species, and not
on ionization state and isotope ratio.
 They imprint well-defined spectral features in the otherwise
featureless continuum spectra of GRBs (and other sources).
This is completely new territory (Iyudin \etal\ 2005), but with the 
great promise to
measure redshifts directly from the gamma-ray spectrum,
i.e. without the need for optical/NIR identification!

Technically, this new strategy requires sensitive spectroscopy
in the 0.2--50 MeV band. The detection of GRBs requires a large
field of view. Therefore, the logical detection principle is a 
Compton telescope.
In addition, such detectors  can be tailored to have
a high polarisation sensitivity. Polarimetry is the last property
of high-energy electromagnetic radiation which has not been utilized
in its full extent, and promises to uniquely determine the emission
processes in GRBs, as well as many other astrophysical sources.
With its large field of view, such a detector will not only
scan 80\% of the sky within one satellite orbital period of 96 min.,
but also provide enormous grasp for measuring the diffuse emission of
nucleosynthesis products and cosmic-ray acceleration.

\section{Scientific Goals}

\subsection{Main mission goal}

We aim at a detection of gamma-ray
bursts at redshifts above 13. This will allow
us to explore rather directly the universe in the epoch where first
stars formed.


\subsubsection{When did the first stars form?}

\noindent{\bf High-redshift GRBs:}
GRB afterglows are bright enough to be used
as pathfinders to the very early universe.
 Since long-duration GRBs are related to the death of massive
stars, it is likely that high-$z$ GRBs exist. Theoretical
predictions range between few up to 50\% of all GRBs being at $z>5$
(Lamb \& Reichart 2001, Bromm \& Loeb 2002),
and stellar evolution models suggest that 50\% of all GRBs occur
at $z>4$ (Yoon \etal\ 2006).
The polarisation data of the Wilkinson
Microwave Anisotropy Probe (WMAP) indicate
a high electron scattering optical depth, hinting
that the first stars formed
in the range 20\lax z \lax 60
(Kogut \etal\ 2003, Bromm \& Loeb 2006, Naoz \& Bromberg 2007).
GRIPS is designed to measure 
GRBs from the death of these first
stars and probe the universe up to the highest redshifts after matter-photon
decoupling.

\noindent{\bf Redshift determination via resonance absorption:}
The observational use of nuclear resonance absorption for GRB redshift
determination depends on two critical questions:
``Is there enough matter along the sight lines
to GRBs?'', and ``Is the resulting absorption detectable?''.

\begin{figure}[h]
\includegraphics[width=.4\textwidth]{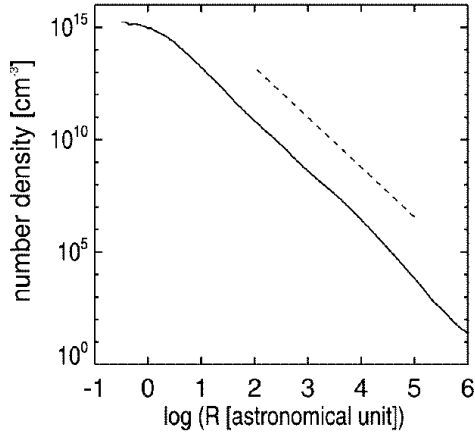}
\hfill\parbox[t]{7.cm}{\vspace*{-2.7cm}\caption[Radial density at z=19]{Radial density around a GRB
progenitor at a redshift z=19. The density profile is close to
a power law $\propto$R$^{-2.2}$ (dashed line). 
It contains about 10$^{28}$ cm$^{-2}$
column density within the inner 1-2 AU, and further
10$^{28}$ cm$^{-2}$ in each shell from 2-10, 10-100, 100-1000 AU!
(From Yoshida \etal\ 2006)
\label{GRBdens}}}
\vspace{-0.4cm}
\end{figure}

 \noindent{\it Is there enough matter along the sight lines to GRBs?}
Apart from galactic foreground extinction, relatively
little intrinsic extinction has been found in the afterglow
sepctral energy distributions
of GRBs, both at X-rays and at optical/NIR wavelengths.
A recent combined analysis of Swift XRT and UVOT
data shows that the
absorbers associated with the GRB host galaxy have column densities
(assuming solar abundances) ranging from (1-8)$\times$10$^{21}$ cm$^{-2}$
(Schady \etal\ 2007).
Yet, there is evidence, both theoretical as well as observational,
that there is substantial amount of matter along the line of sights
to GRBs. This applies to the local GRB surrounding as well as to
the larger environment of the host galaxy in which the GRB explodes.
Temporally variable optical absorptions lines of fine-structure
transitions indicate that (i) all material at distances within
a few kpc is ionized, most likely by the strong UV photon 
flux accompanied with the emission front of the GRB,
and (ii) beyond this ionized region the absorbing column 
is still at a level of 10$^{21}$ cm$^{-2}$  (Vreeswijk \etal\ 2007).
Thus, present-day measurement capabilities in the optical/NIR as well
as X-rays are not adequate to determine the density of local matter 
around GRBs.
However, at $\gamma$-rays this matter will be measurable 
through nuclear resonance absorption even though this matter is 
simultaneously being ionized: the GRB gamma-ray radiation
has to pass through it - and it will suffer resonance absorption independent
on whether this material is ionized or not.

A variety of theoretical simulations of
GRB progenitors have been made (e.g.
Bate \& Bonnell 2003, Yoshida \etal\ 2006, Abel \etal\ 2007, Gao \etal\ 2007),
pertaining to
the formation of the first stars, the fragmentation rate,
and density structure around the first star.
The first stars are thought to form inside halos of mas
10$^5$...10$^6 M_{\odot}$ at redshifts 10-60.
It is generally accepted that most of the 10$^5$...10$^6 M_{\odot}$ halo mass 
remains in the surroundings of the forming proto-star,
with about the original dimensions of the proto-cloud. 
The resulting mass of the star as well as the density
structure are difficult to predict because they depend on the
collapse conditions (merger or not, strength of winds, etc).
However, it is important to realize that some simulations in fact predict
column densities of up to 10$^{29}$ cm$^{-2}$ around the first stars
(Yoshida \etal\ 2006,  Spolyar \etal\ 2008;  see also Fig. \ref{GRBdens})! 
These simulations have been done independent
of the knowledge of nuclear resonance absorption.
It remains to be demonstrated (preferentially observationally) whether 
the conditions modelled in these simulations are realized.
Yet, the existence of what one ``normally''
would refer to as ``unbelievably high'' column densities is plausible -- note 
that even pristine and fully-ionized hydrogen imprints resonant absorption!
GRBs are the best and possibly only tool to measure such conditions.

\begin{figure}[th]
\includegraphics[bb=77 366 558 692,width=7.3cm,clip]{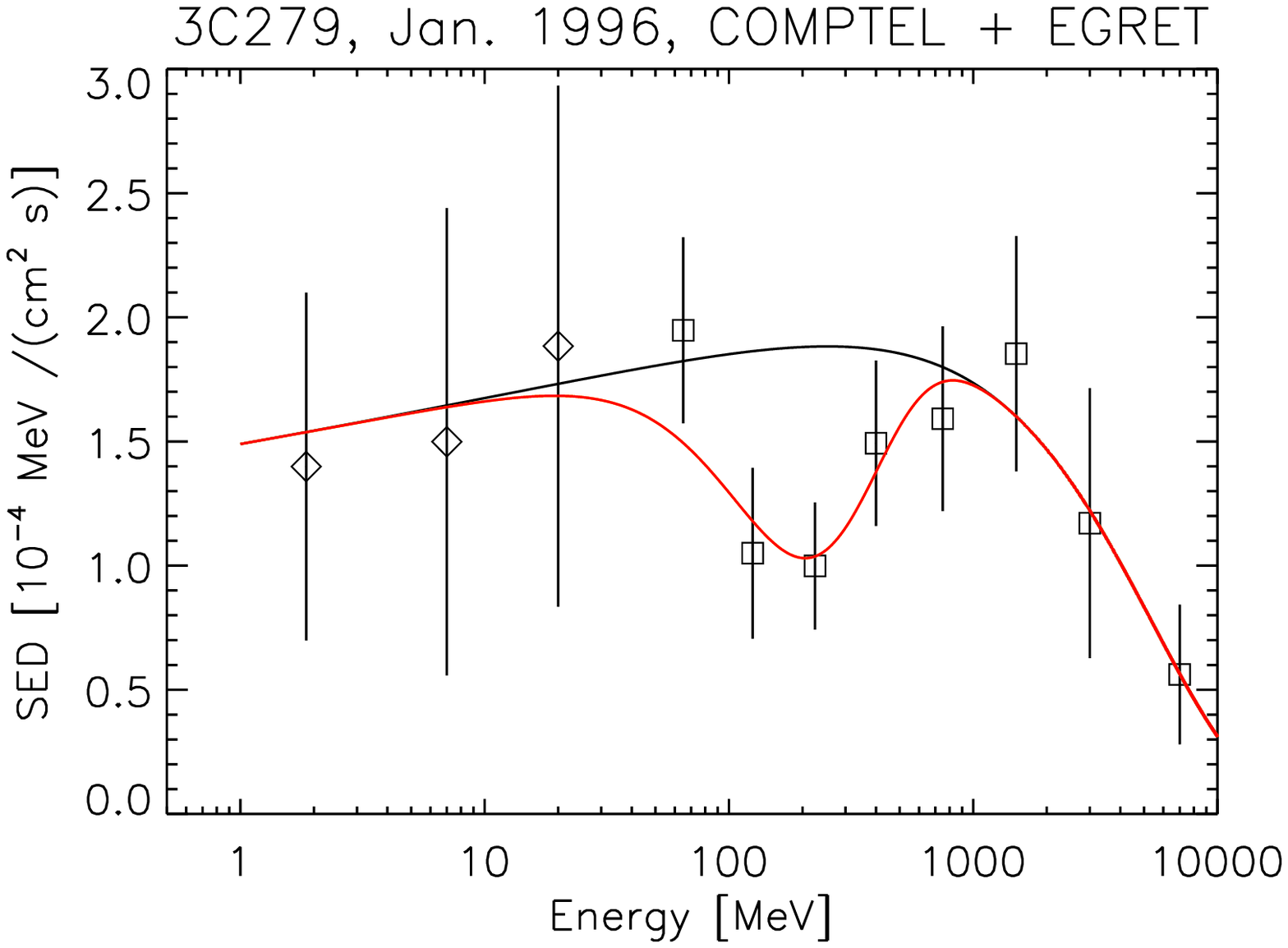}
\includegraphics[width=7.5cm]{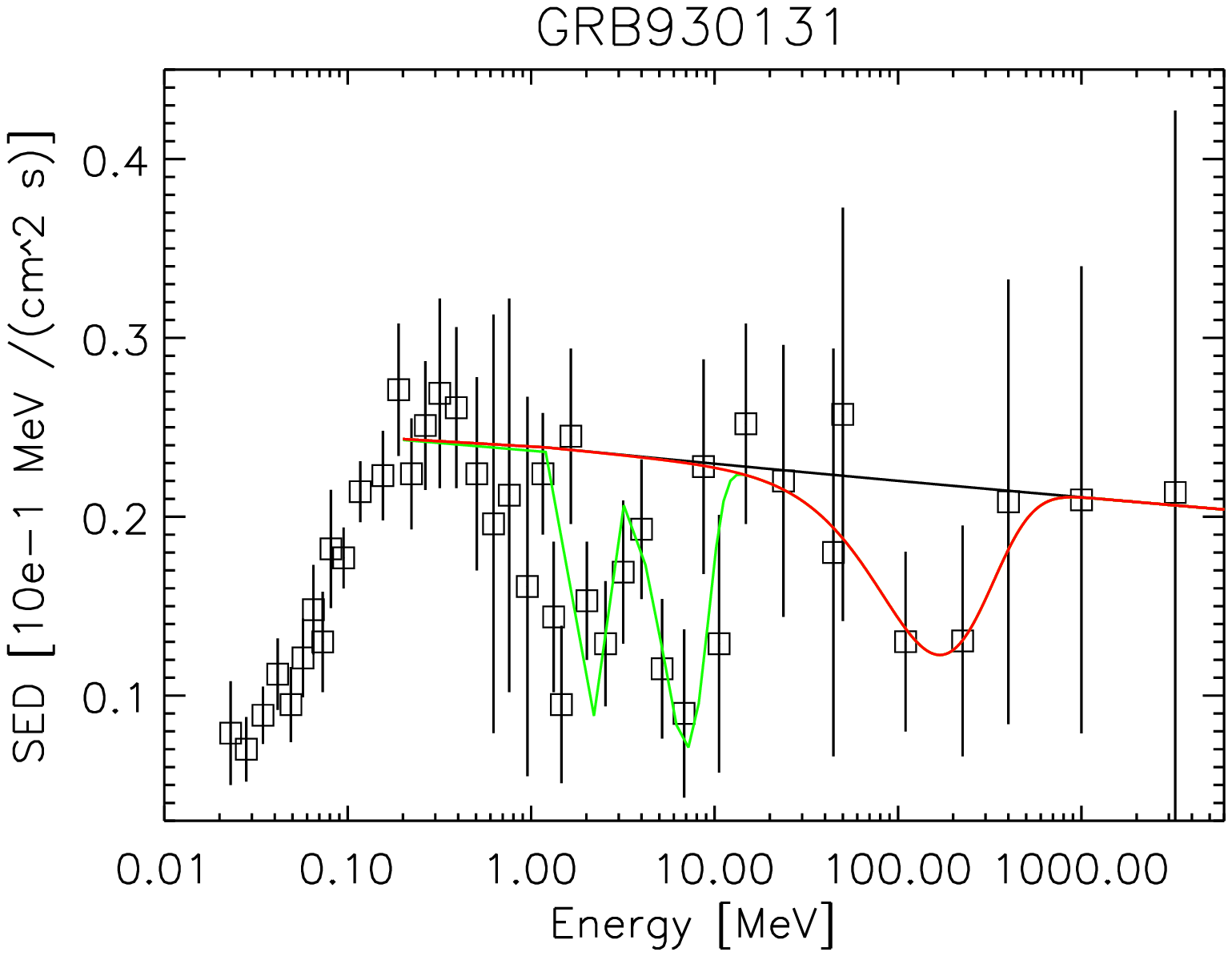}
\vspace{-0.3cm}
 \caption[Delta resonance in 3C 279 and GRB 930131]{{\bf Left:} 
3C 279 spectrum during the January
1996 flare as measured by COMPTEL (diamonds) and EGRET (squares),
which includes the ${\Delta}$ resonance
absorption in the circumnuclear environment (red line) in addition to
a cut-off power law (black line).
The best-fit energy for the ${\Delta}$ resonance is 208$\pm$25 MeV,
implying a redshift 0.57 $\pm$ 0.12, close to the
optically determined z=0.536.
{\bf Right:} 
Fit to the combined
COMPTEL/EGRET spectrum of GRB 930131. The throughs at 5--8 MeV and
100--200 MeV are
compatible with the Giant Dipole and Delta resonance, respectively,
 at a redshift of z$\sim$1 (Iyudin \etal\ 2007b).
\label{grb93}}
\end{figure}

\noindent{\it Is the resulting absorption detectable?}
This question can be subdivided into two issues: Firstly,
are there astrophysical
conditions which favour large line of sight columns, but
not excessively large densities? If the density is too high,
multiple-scattering of higher energy photons could partly
fill the energy window of a resonance, thus smearing the absorption trough.
This may happen via Compton scattering or
 the cascading of high-energy photons. 
While the total pair production or Compton scattering 
cross sections are about a factor of 30-40 larger than the 
peak cross section of the Giant Dipole or Delta Resonance,
the jet geometry in both, GRBs as well as blazars, over-compensates
for this statistical measure: it is the differential cross section
which matters.
At the Dipole Resonance energy,
the Compton-scattered photon beam has a full-width-half-maximum
of 16\degs, or 0.5 sr. For a 1\degs\ opening angle of the jet,
the resulting GDR absorption is a factor $\sim$12 more efficient
than the Compton re-scattering of higher energy continuum photons
into the beam. In addition, Compton scattering changes the energy
of the scattered photon by arbitrary large values, much greater 
than the width of the resonance - this adds another factor
of E/$\Delta$E (\gax 3 for the GDR) in favor of the resonance
absorption. For pair production, only the latter effect comes into play.
Furthermore,  even in high-density
environments, there are two possibilities which allevate the problem
of re-filling:
(i) the transverse dimension of the absorber is less
than $\sim$1.5 attenuation lengths at the energy of the highest
attenuation value (Varier \etal\ 1986) or (ii) the absorber consists
of many clumps (clouds) of matter,
a solution which has been proposed to explain the UV and X-ray
(Arav \etal\ 2003; 2005) or IR emission (Elitzur \etal\ 2004)
in optically thick absorbers around AGN.
First hints for resonance absorption in AGN
(Fig. \ref{grb93}, left panel; Iyudin \etal\ 2005), and GRBs 
(Fig. \ref{grb93}, right panel)
at the
2$\sigma$ level have been found based on COMPTEL and EGRET data.

\begin{figure}[ht]
\includegraphics[width=.5\textwidth, angle=0]{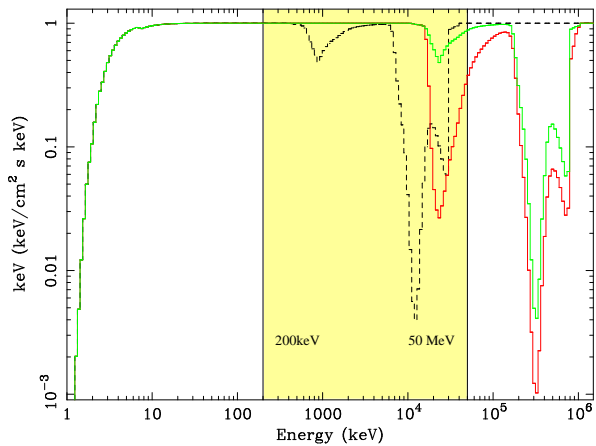}
\hfill\parbox[t]{6.5cm}{\vspace*{-5.5cm}\caption[Resonance absorption at different Z and z]{Resonance
absorption lines for two different redshifts (black: z=25; color: z=0)
and different metallicities: Z=0.1 (green) and Z=1 (red) solar metallicity.
Note the obvious difference in the relative strengths of the absorptions.
The solid vertical lines bracket the energy band of GRIPS from 200 keV
to 50 MeV. A total column of 10$^{28}$ cm$^{-2}$ has been assumed to clearly
visualise the effect, and a density smaller than 10$^5$  cm$^{-3}$
to avoid refilling of the lines due to scattering. The detectability
of these resonance lines with the proposed instrument is achieved 
down to column densities of 10$^{25}$ cm$^{-2}$ (Fig. \ref{resores}).
\label{reso}}}
\vspace{-0.4cm}
\end{figure}

Secondly, how narrow/broad will the
resonant absorption features be at different redshifts, and what is the
required energy resolution and sensitivity to detect them?
Fig. \ref{reso} shows the rest-frame resonance absorption lines for
two different metallicities which clearly illustrate their
starkly different relative intensity ratios. The GDR around 25 MeV
has a FWHM $\sim$ 10 MeV, thus GRIPS' $\sim$300 keV resolution
at 10 MeV is fully sufficient
to properly resolve the feature. The goal of GRIPS is to measure
those resonance absorptions at redshifts $z \sim 13-30$.
A simulated GRB spectrum at z=25 is also shown in  Fig. \ref{reso}.
Full simulations for the $\gamma$-ray instrument on GRIPS 
show that column densities of 10$^{26}$...10$^{27}$ cm$^{-2}$ will be
detectable for $\sim$ 30--40\% of all GRBs seen in spectroscopy mode,
and those with as little as 10$^{25}$ cm$^{-2}$ for the 10\% brightest GRBs
(see sect. 4.2.4).

\begin{figure}[ht]
\centering{
\includegraphics[width=0.45\columnwidth]{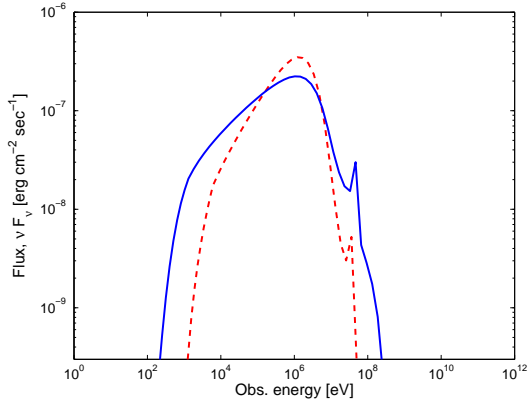}}
\hfill\parbox[t]{7.cm}{\vspace*{-4.cm}\caption[Theoretical GRB spectra with annihilation]{Spectra averaged over the GRB duration,
obtained for high compactness $l$. The two models assume
 $\Gamma$=300,  L = 10$^{52}$ erg,  z = 1, and
differ in the variability timescale of $\delta$t = 10$^{-4}$ s, 
(blue solid line; implied l=250);
and $\delta$t = 10$^{-5}$ s (dashed line; l=2500)
at which the Lorentz factor $\Gamma$ varies.
The scattering optical depth at the
end of the dynamical time is 13 and 56, respectively.
The peaks observed  at $\sim$80 MeV result from pair annihilation
which is larger for larger compactness.
(From Pe'er \& Waxmann 2004)}
\label{grbanni}}
\end{figure}

These features have not been seen before in EGRET TASC, BATSE SD, or
COMPTEL (D2)
data due to the combination of insufficient sensitivity, poor energy
resolution and pre-canned response matrices for fixed energy bins.
GRIPS will move $\gamma$-ray astronomy into a new era combining sensitivity,
energy resolution and a large FOV in one instrument that will
detect nuclear resonance troughs in the spectra of a large number of
sources.

If the GRB environment contains total (neutral plus ionized)
column densities of 10$^{25}$ cm$^{-2}$ or higher,
GRIPS will be able to measure redshifts via the nuclear resonance
absorption directly from the $\gamma$-ray spectrum, and thus be sensitive
well beyond z$\sim$13.

\subsubsection{What are the energetics of GRBs?}

There is increasing evidence, both observational (Starling \etal\ 2005)
and from theory (Woosley \etal\ 2006) that a GRB
may launch two jets: one highly relativistic ($\Gamma > 200$, kinetic energy
of 10$^{51}$ erg), central jet with
an opening angle of a few degrees, and a broader jet with sub-relativistic
ejecta (0.1c, kinetic energy 10$^{52}$ erg) spread over 1 radian.
The underlying physical basis of the correlation between the isotropic 
equivalent energy emitted in the
prompt radiation phase and the peak energy (E$_{\rm peak}$) of the measured
spectrum (Amati \etal\ 2002,
Ghirlanda \etal\ 2004)
is mysterious.
Measuring the Lorentz factor $\Gamma$ from the $\gamma$-ray emission itself
with GRIPS would
provide a major step in understanding the cause of this correlation
and in determining the energetics of GRBs.

Numerical models of the prompt emission due to internal shocks
in an expanding relativistic wind exist in many different variants.
In most of these, the spectrum is expected to
significantly deviate from an optically thin synchrotron spectrum.
A time-dependent calculation which includes cyclo-synchrotron emission and
absorption, inverse and direct Compton scattering, and e$^+$e$^-$ pair
production and annihilation has shown (Pe'er \& Waxman 2004)
that (i) (in-flight) annihilation emission lines
are expected at surprising strength (Fig. \ref{grbanni}),
and (ii) thermal Comptonization
leads to emission peaking at \gax30 MeV, possibly explaining the
additional MeV component seen in GRB 941017 (Gonz\'alez \etal\ 2003).
The annihilation line will be boosted by $\Gamma$ into the \gax 100 MeV range,
but adiabatic energy losses as well as GRB
cosmological redshift lowers the observed photon energy back into the few
MeV band.
Thus, once the redshift is determined by the resonance lines (or
ground-based optical/NIR telescopes for z$<$13),
the measured energy of the annihilation line allows a direct
determination of the Lorentz factor from the prompt gamma-ray emission
spectrum. 
GRIPS will allow us to
measure the Lorentz factor of the prompt $\gamma$-ray burst emission
via the predicted annihilation line,
and thus directly measure the total energy of the explosion
for at least 15\% of all GRBs.
Since the measured annihilation line energy is proportional to the ratio
$\Gamma$/$z$, GRIPS can detect the line in the range
(0.5 \gax $\Gamma$/(1 + $z$) \gax 100), thus covering
the full predicted range of $\Gamma$ (50--1000) and $z$  (0.5-60). 

\subsubsection{What is the emission mechanism?}

\noindent{\bf Spectra:}
Based on the knowledge gained from observation in the keV-MeV
range several possible radiation mechanisms exist, all of which
produce characteristic spectra in the super-MeV range. Two
main classes of models  (see
M\'esz\'aros 2006) have been discussed. (1) Synchrotron/inverse Compton
emission of electrons and protons: It is very probable that
particles are accelerated to very high energies close to the
emission site. This could either be in shock waves, where the
Fermi mechanisms or other plasma instabilities  act, or in
magnetic reconnection sites. Therefore, it is likely that the
observed emission in the keV range scatters on these relativistic
electrons, which will result in inverse Compton emission in the
super-MeV domain.  This occurs in both internal and
external shock scenarios. Furthermore, most of the outflow
energy, transported by the protons
will predominantly be in the super-MeV range.
(2) Hadron related emission via pion production and
cascades: High-energy, neutral pions ($\pi^0$) can be created as
shock-accelerated, relativistic protons scatter inelastically off
ambient photons ($p\gamma$ interactions). These later decay into
$\gamma$-rays. This is, e.g., suggested to occur in 
 GRB 941017 (Gonz\'alez \etal\ 2003). Similarly, if the
neutrons in the outflow decouple from protons, inelastic
collisions between neutrons and protons can produce pions
and subsequent high-energy emission.


The spectra of some GRBs alternatively have been well fit by both the Band
model and a combination of a black body plus power law model
(McBreen \etal\ 2006, Ryde \etal\ 2005).
Rees (2005) suggested that the
$E_{peak}$ in the $\gamma$--ray spectrum is due to a Comptonised
thermal component from the photosphere, where the comoving optical
depth falls to unity. The thermal emission from a laminar jet when
viewed head--on would give rise to a thermal spectrum peaking in
the X-ray or $\gamma$-ray band. The resulting spectrum would be
the superposition of the Comptonised thermal component and the
power law from synchrotron emission.
Thus, from theory there is no indication that GRB spectra should
deviate from a smooth continuum,
and thus can be used as reliable background light sources for nuclear
absorption features.

GRIPS will measure the broad-band spectra of $\sim$660 GRBs/yr
from 100 keV -- 50 MeV with 3\% energy resolution.
 GRIPS will distinguish between the various, contradictory
mechanisms  by covering the transition from the classical keV and the
dozen MeV regime.

\noindent{\bf Polarisation as a diagnostic of the GRB emission mechanism:}
The link between the
$\gamma$-ray production mechanism in GRBs and the degree of linear
polarisation can constrain models.
A significant level of polarisation can be produced in GRBs by
either synchrotron emission or by inverse Compton scattering. The
fractional polarisation produced by synchrotron emission in a
perfectly aligned magnetic field can be as high as
70--75\%. An ordered magnetic
field of this type would not be produced in shocks but could be
advected from the central engine
 (Granot \& K\"onigl 2003, Granot 2003,  Lyutikov \etal\ 2003).
 It should be possible to distinguish between
Synchrotron radiation from an ordered magnetic field advected from
the central engine and Compton Drag. Only a small
fraction of GRBs should be highly polarised from Compton Drag
because they have narrower jets, whereas the synchrotron radiation
from an ordered magnetic field should be a general feature of all GRBs.

GRIPS will allow us to measure the polarisation of the prompt
$\gamma$-ray burst emission to a few percent accuracy for about 10\% of all
detected GRBs. Moreover, the superior polarisation sensitivity
will even securely measure whether
or not the percentage polarisation varies with energy, angle and/or time
over the burst duration of a dozen brightest GRBs.

\subsection{Other mission goals}


GRIPS will address other,
non-GRB science topics, and lead to
guaranteed science returns. Major science topics are (a) to unveil the
physics of stellar explosions, and (b) to illuminate the physical
processes which lead to particle accelerations from thermal up to
relativistic (cosmic-ray) energies.

With the imminent launch of NASA's GLAST Observatory (30 MeV - 100 GeV)
and the long-term continuation of ESA's INTEGRAL Observatory (20 keV - few
MeV) there will be a dramatic lack of coverage of the gamma-ray sky in
the
1-30 MeV range. GLAST will detect thousands of new sources, while the
only
existing MeV data (from GRO/COMPTEL) is for a handful of very bright
ones.
The discovery potential is clearly enormous. At present, multiwavelength
spectra of the majority of sources have a notable absence of data at MeV
energies, although often the main power is expected there. 
Without a mission like GRIPS, the MeV sky with its wealth of astrophysics will
be less well known even
than the TeV sky (via the new Cerenkov observatories), which 
is a quite disillusioning prospect.

\begin{figure}[hb]
\includegraphics[width=0.33\columnwidth]{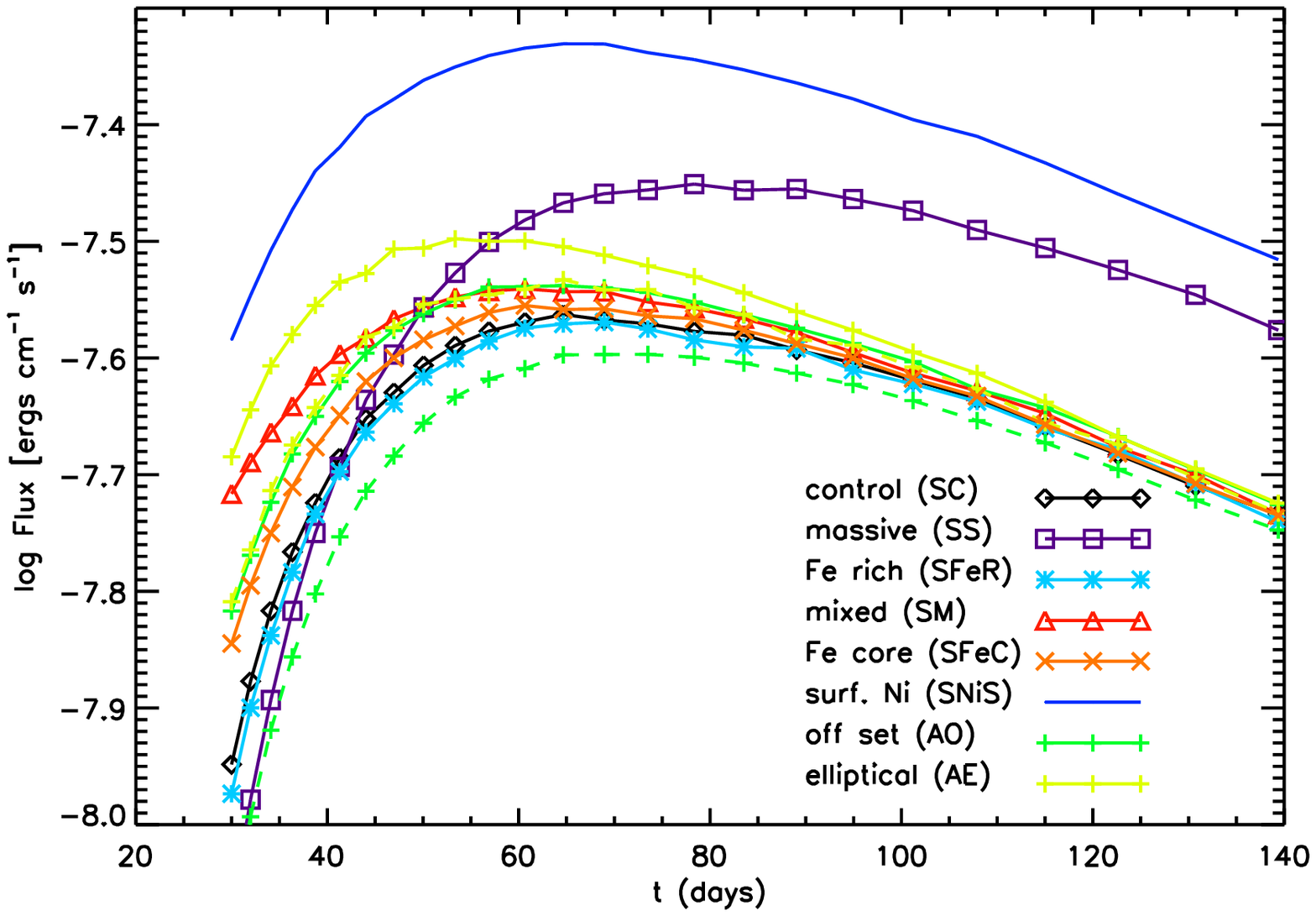}
\includegraphics[width=0.33\columnwidth]{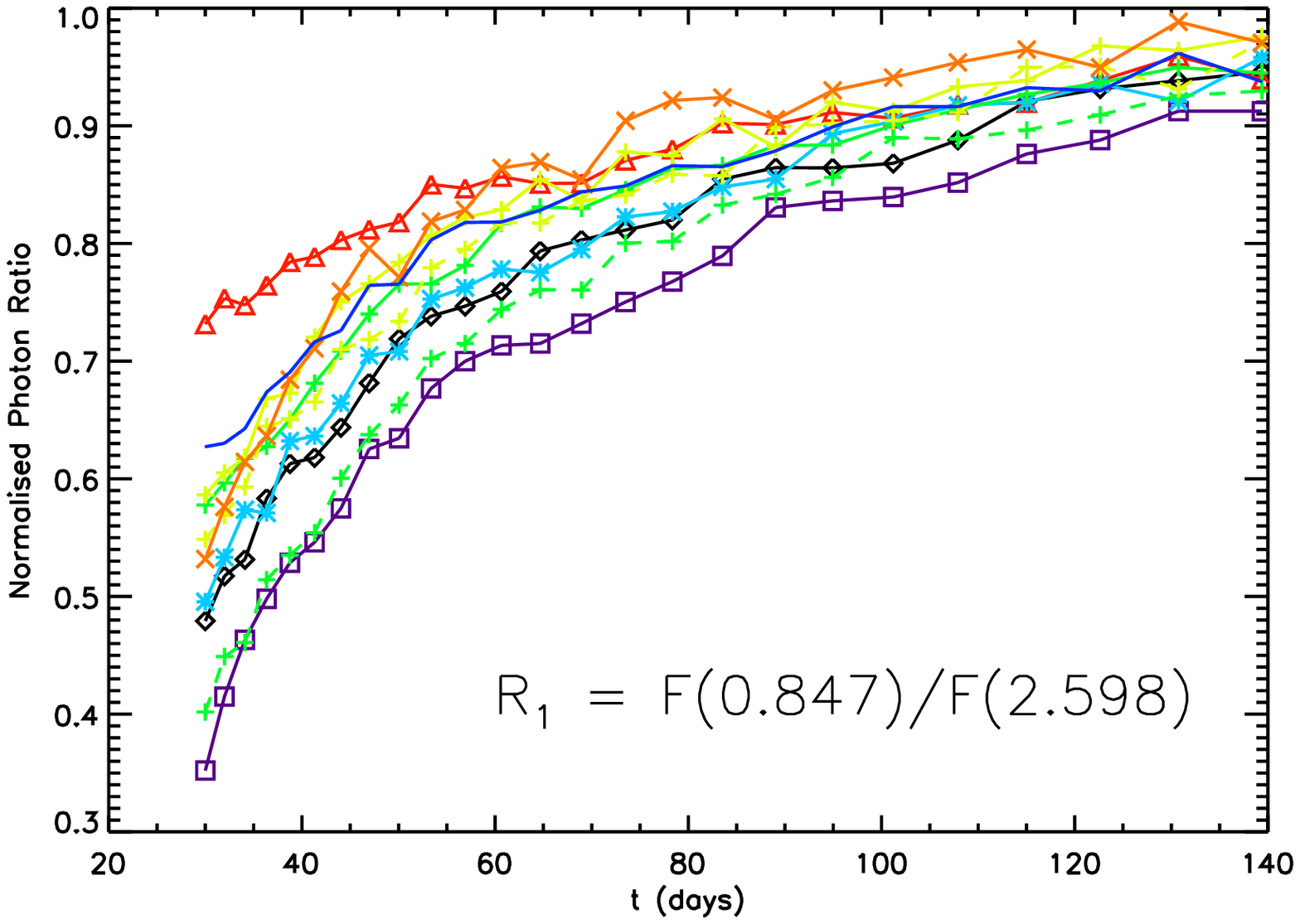}
\includegraphics[width=0.33\columnwidth]{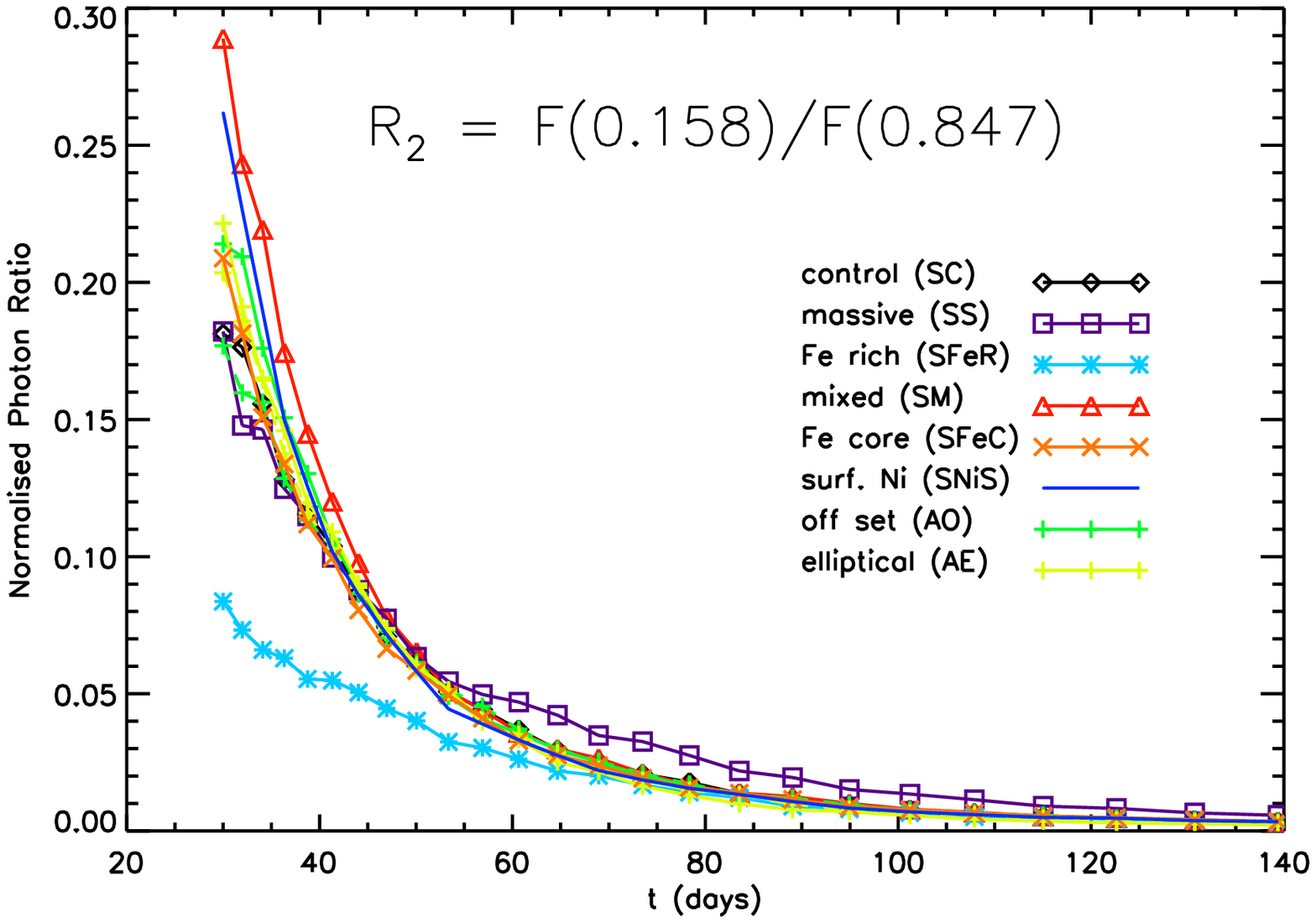}
\caption[Line ratio in SNIa]{SNIa model light curves in the 0.1-3.5 MeV band
for a 1 Mpc distance, and time since explosion (left), as well as
ratios of gamma-ray lines from $^{56}$Ni
decay (middle and right) which
provide an important
diagnostics of the inner supernova and the explosion mechanism.
Shown are spherically-symmetric models as well as two cases of
asphericities (AO and AE) which represent the extremes with 
respect to viewing angle.
The high-energy lines 
provide a sensitive diagnostics
to asphericities of the SN. (From Sim \& Mazzali 2008)
\label{fig_sn-lineratio}}
\end{figure}

\subsubsection{Nuclear astrophysics}

{\bf Supernovae of Type Ia} (SNIa) are considered on purely empirical
grounds to be standardizable candles; no physical explanation could be
established for the homogeneity of thermonuclear disruptions of white
dwarf stars, which is the widely-accepted model. In view of the
far-reaching cosmological implications of the apparent dimming of
SNIa in the distant universe, this remains a major concern for supernova
scientists (Leibundgut 2001, Branch \& Nomoto 2007).
Phrased at an extreme: dark energy might not exist at all, if our
estimates of SNIa properties across cosmic times  or selection biases
are inadequate!   Already a 7\% Ibc contamination level is sufficient
to produce $\Omega_{\lambda}$ = 0.7 from no effect (Homeier 2005).
Thus, understanding the physics of SNIa is of utmost importance.
GRIPS will provide the data for a
    physical understanding of the Phillips relation and the related errors.
Due to the huge dynamic scales in time and space of the relevant physical
processes, numerical simulations must make use of approximations.
Guidance from observations is essential in building
a physical model. 
Most direct access to isotope information would be through
nuclear lines emitted from radioactive decays, which even can provide
kinematic information
from partially-embedded freshly-synthesized species.
Penetrating $\gamma$-rays are expected to escape from the supernova
as early as a few days after the explosion.
$^{56}$Ni decays to $^{56}$Co within $\sim$8 days, which then decays to
stable $^{56}$Fe within  $\sim$111 days, producing $\gamma$-ray lines at
158~keV ($^{56}$Ni, early and probably occulted)
and at 847, 1238, 1771 and 2598~keV.
Line shape and centroids reflect the
original $^{56}$Ni kinematics, line ratios  are a key diagnostic
of the explosion morphology and hence model types.
 In the period before full  $\gamma$-ray
transparency is reached ($\sim$100 days),  $\gamma$-rays
from $^{56}$Co decay provide the key information; the GRIPS band is
designed to include these lines.
Recent SNIa surveys record $>$10 nearby events ($<$50 Mpc;
Isern, priv.comm.), 
which will be detected by GRIPS in the energy range of the (direct and 
scattered) $^{56}$Ni decay $\gamma$-ray range up to $\sim$2.5 MeV at $\sim$100 
days after explosion, when the SN is transparent to gamma-rays 
(see Sim \& Mazzali 2008). In addition to an absolute determination of 
the $^{56}$Ni amount from the $\gamma$-ray line flux, due to its broad energy 
range, line ratio diagnostics will be provided by GRIPS, thus significantly 
enhancing the sample of SNIa with meaningful $\gamma$-ray constraints.

GRIPS will at least measure one SNIa per year in one or more of the $^{56}$Ni 
decay gamma-ray lines, to constrain explosion models through energy-dependent 
transparency and the absolute $^{56}$Ni radioactivity.

{\bf Massive-star nucleosynthesis} is responsible for most of the
intermediate-mass elements from oxygen through iron (Woosley \& Weaver 1995;
Heger \etal\ 2003). Different burning episodes from hydrogen
through silicon burning in shells of the rapidly-evolving star after its
core-hydrogen burning phase (main sequence), plus nucleosynthesis in the
supernova following the final gravitational collapse of the star, are
responsible for this element production. The complexities of stellar-structure
evolution and active nuclear-reaction networks are difficult to model. 
Beyond the (precise and abundant, but rather
indirect) observations of elemental abundances through atomic
lines, measurements of the key isotopes through
their radioactive decays provide calibrators of those models.

A key isotope for supernova-interior nucleosynthesis is $^{44}$Ti with a decay
time of $\sim$85 years, a  $\gamma$-ray line at 1156 keV and X-ray lines
at 68 and 78 keV.  This radioactive
decay has been observed from the Cas A supernova remnant
(Iyudin \etal\ 1994, Vink \etal\ 2001).
Its abundance and
kinematics  directly arises from the processes of accretion
and fall-back onto the central remnant.
This otherwise unaccessible inner
region of a core collapse
is at the origin of gamma-ray burst formation by massive stars (The et
al. 2006). 
GRIPS will deepen surveys for $^{44}$Ti gamma-ray sources (1.16 MeV) in the 
Galaxy within a 5-year mission down to 
6$\times$10$^{-7}$ ph cm$^{-2}$ s$^{-1}$,
and should detect 10-15 sources if $^{44}$Ti ejection is typical for 
core-collapse supernovae (see The et al. 2006).
Furthermore, detection of MeV continuum in core-collapse SNe, and
especially in the Ib/c class, would
indicate that a fraction of the relevant kinetic energies liberated in
these explosions is conveyed to the
acceleration of electrons to very high energies and to the re-emission
through the synchrotron process,
illuminating our understanding of the GRB-powering mechanisms in SNe.
Moreover, if the high energy
spectrum were polarized, we would get unique insights into the geometry
of the magnetic fields.
For a sufficiently closeby (less than 100 Mpc) SN associated with a GRB,
prompt gamma-ray
spectroscopy of possible early $^{56}$Ni lines at 158 keV and 812 keV would
be another unique opportunity, according to
the predicted gamma-ray flux near the time of 
transparency ($\sim$50-100 days) (see Sim \& Mazzali 2008). 
GRIPS would be the only monitor for such continuum and radioactivity MeV
gamma-rays from nearby
core-collapse supernovae, with these unique signatures.


The $^{60}$Fe isotope is produced in late burning stages of massive stars,
and only ejected by the terminal supernova.
$^{60}$Fe decays with  $\sim$2.2 My, hence cumulative
production over My time scales is observed in the  $\gamma$-ray lines.
INTEGRAL detected the $\gamma$-ray lines
from this isotope in the larger region of the inner Galaxy (Wang \etal\ 2007).
Yet, spatial information is insufficient to be able to conclude on an
origin from core collapse SN.
With a 5-year survey sensitivity of 
$\sim$6$\times$10$^{-7}$ ph cm$^{-2}$ s${-1}$, GRIPS will map the Galaxy 
in $^{60}$Fe radioactivity gamma-rays for the first time at several degrees 
spatial resolution. This will allow detailed tests of massive-star 
nucleosynthesis models for massive-star regions of different ages along the 
plane of the Galaxy.

The $^{26}$Al isotope is predominantly produced by massive stars
(Prantzos \& Diehl 1996), although not only by supernovae, but
with nucleosynthesis in early stages
of core hydrogen burning and in late shell burning stages plus
supernova nucleosynthesis (Limongi \& Chieffi 2006). With a 1.04 My
decay time, its cumulative emission in the 1808.65 keV $\gamma$-ray line
provides a diagnostic of massive-star nucleosynthesis in the Galaxy
(Diehl \etal\ 2006). The $^{26}$Al production in
nearby massive-star clusters with correspondingly-low surface brightness
are within reach with the sensitivity of GRIPS.
This is a key ingredient of chemical-evolution models of
galaxies
(MacLow \& Klessen 2004).
GRIPS will deepen the Galactic surveys for these key isotopes of
massive-star nucleosynthesis, and provide significant improvement on the
Galactic distribution of massive-star  radioactivities.

{\bf Novae}
are understood as explosive Hydrogen burning on the
surface of a white dwarf (WD), igniting once a sufficient amount of matter
has been accreted from the WD's companion star.
Nuclear burning is dominated by rapid proton capture (rp-process) on
light elements, and the main theoretical issue
is up to which heavy isotopes the rp-process proceeds.
Many proton-rich isotopes
are produced, which undergo $\beta^+$-decay at their characteristic decay
times, most important contributions being from $^{13}$N ($\tau\sim$14~min)
and $^{18}$F ($\tau\sim$2.5~h) (Hernanz \etal\ 2002, Hernanz \& Jose 2006).
Therefore, a bright flash of positron annihilation emission with a
characteristic
line at 511~keV and a bright lower-energy continuum occurs right
after the thermonuclear runaway, well before the envelope expands and
lights up as a nova in optical emission.
Alongside the rp-process, radioactive $^7$Be is produced and is
expected to be the brightest $\gamma$-ray line source, with a decay
time of 78~days (E$_{\gamma}$=478~keV).

GRIPS will detect several Galactic novae through annihilation
emission, and combine positron annihilation with $^7$Be and $^{22}$Na
diagnostics to understand the nova ignition and burning process.

\subsubsection{Positron Astrophysics}

\vspace{-0.2cm}

Imaging of the  $\gamma$-ray emission from the annihilation of positrons
with INTEGRAL/SPI (Kn\"odlseder \etal\ 2005)
has revealed a new major puzzle:
The expected sources of positrons in the Galaxy are
 predominantly located in the disk of the Galaxy, while the Galactic
bulge is by far the brightest observed feature in annihilation $\gamma$-rays.
This initiated a quest for new
types of candidate positron producers, such as the annihilation of dark matter
particles accumulated by the Galaxy's gravitational field
(Hooper \& Wang 2004).
An alternative
hypothesis could be the large-scale transport of positrons from their
disk sources through the Galactic
 halo into the central bulge before their annihilation (Prantzos 2004), or
significant diffusion of positrons generated by past activity and energetic
proton ejection from the central black hole in our Galaxy (Cheng \etal\ 2006).

GRIPS will help to clarify the validity of such extreme models, as it
will be sensitive to the full spectral range of annihilation emission
(~300$\rightarrow$511~keV, and above, for annihilations-in-flight).
With its angular resolution of $\sim$3\fdg 5 at 511 keV, GRIPS will map the 
bright bulge emission of positron annihilation and its merging with the disk 
component; the latter can be mapped with GRIPS along the bright regions of the 
disk such as the inner ridge and Cygnus, as inferred from $^{26}$Al gamma-rays 
and its associated positron production (see Zoglauer \etal\ 2008).

\subsubsection{Other $\gamma$-ray sources}

{\bf Blazars:}
Recent Swift/BAT observations find many blazars with very flat (photon
index 1.6--1.8) spectra in the 20--150 keV range, and extrapolations
indicate that all of those will be securely detectable by GRIPS.
The measurement of the resonance absorption
will substantially help in the identification process, since the redshift
will be known from the gamma-ray spectra (Iyudin \etal\ 2005, 2007a).
Gamma-ray polarization can be measured for the flare states.
Comparing with radio polarization and multi-wavelength
light-curves, this will pin down the origin of the $\gamma$-ray flares
(Wehrle \etal\ 2001;
Jorstad \etal\ 2006).

\noindent{\bf Pulsars, AXPs, and SGRs:}
Pulsars
 are an excellent
laboratory for the study of particle acceleration,
radiation processes
and fundamental physics in environments characterized by strong gravity,
strong magnetic fields, high densities and relativity.
Above $\sim 50$ MeV EGRET has detected 6-10 pulsars and candidates
(Thompson \etal\ 1999) and Sch\"onfelder \etal\ (2000)
reported for the COMPTEL survey 4-5 pulsars corresponding to EGRET
detections and one high-magnetic field pulsar (PSR B1509-58) with
emission up to 30 MeV.
Estimates for the number of possible pulsar detections by future, more
sensitive, gamma-ray instruments are based on our empirical knowledge
of pulsar efficiencies and spectra, new radio surveys that now contain
about 2000 pulsar detections, and on the extrapolations afforded by
theories of high-energy emission from rotating, magnetized neutron stars.
These theories are still quite disparate. For energies above several 10
MeV and for the sensitivity of the Large Area Telescope on GLAST,
predictions range from $\sim 60$ detections (Harding \etal\ 2007) to
about 800 (Jiang \& Zhang 2006), with several predictions between 120
and 260 pulsars. Many of these pulsars are so-called 'radio-quiet'
objects comparable to the Geminga pulsar. Additional predictions for
about 100 millisecond-pulsars have been published for GLAST by Story 
\etal\ (2007). GRIPS' sensitivity for wide band spectra above $\sim 10$ MeV
(normal pulsars) and below 1 MeV (high B-field pulsars and AXPs) exceeds
the COMPTEL sensitivity by a similar factor that holds when going from
EGRET to GLAST. From the EGRET/COMPTEL relation we therefore estimate
the number of GRIPS pulsars to be about 50-70\% of the GLAST pulsars and
expect to detect 60-70 pulsars in the GRIPS energy range $>10$ MeV.
GRIPS-determined light curves and phase-resolved spectra
in the 1--50 MeV range will provide
decisive insights into the pulsar magnetosphere and the
acceleration processes located there.
Polarisation is a unique
characteristic of particles radiating in strong magnetic fields.
Pulsars and their surrounding pulsar wind nebulae (PWN) are highly
polarised.
GRIPS will be sensitive enough to measure polarisation below a few MeV
from several pulsars.

\noindent 'Magnetars'
appear to observers in the forms of 'anomalous X-ray
Pulsars' (AXP) or, possibly related,  'soft gamma-ray repeaters'
(SGR).
 Pulsed radiation with a thermal spectrum at soft X-rays
(1--10 keV) and an extremely hard power-law up to nearly 1 MeV has been
observed from 6-8 AXPs (Kuiper \etal\ 2006). 
The continuous all-sky survey of GRIPS promises to capture unique data for
high-energy neutron star astrophysics.

\noindent{\bf Superbubbles:}
Massive stars in the Galaxy appear in groups
(e.g. OB associations), such that their strong wind and supernova activity
overlaps and generates large  superbubbles (SBs).
Interacting shocks within the superbubble
thus provides a natural environment for cosmic ray accerelation
(GCRs).
GRIPS has sufficient
sensitivity to test the theory of galactic cosmic ray acceleration in the
SB environment by observing galactic SBs as well as
 30 Dorado in the LMC in gamma-ray line emission.
Lines of $^{12}$C$^*$ and of $^{16}$O$^*$ at 4.44 MeV,
and 6.1 MeV, 6.97 MeV, and 7.17 MeV, respectively, would be the best
indicators of the CR acceleration in the SB.


\noindent{\bf Diffuse continuum MeV gamma-ray emission}
from the interstellar medium arises from
the interactions of cosmic-ray electrons with interstellar matter and
radiation fields.
Bremsstrahlung on atomic and molecular hydrogen produces gamma-rays
with energies
typically half of the electron energy, so that gamma-rays in the GRIPS
range  trace
electrons in the 1-100 MeV ranges.
At these energies direct cosmic-ray measurements are virtually impossible
because of the large solar modulation; synchrotron radiation occurs at
frequencies too low to be observable.
Hence GRIPS gamma-ray observations are our only window to interstellar 
electrons at MeV energies.
Inverse-Compton scattering of 100-1000 MeV
 electrons and positrons on interstellar radiation is an important
 component of continuum gamma-rays, and dominates at
 low energies. In fact, recently this component has been shown to explain
most of the diffuse emission from the Galactic ridge observed by INTEGRAL
(Porter et al. 2008).
Both primary and secondary cosmic-ray electrons and positrons  contribute to
this emission, and hence observations in this part of the spectrum give
valuable information on high-energy particles in the Galaxy.
 On the other hand
this process fails to explain all of the diffuse emission observed
by COMPTEL (Strong et al. 2005, Porter et al. 2008),
suggesting there could be a contribution from populations of unresolved
hard gamma-ray sources, like AXPs and radio pulsars.
GRIPS can trace and disentangle those
components towards
higher energies.



{\bf Solar flares} accelerate ions to tens of GeV and electrons to tens of
MeV. 
GRIPS would obtain high-statistics time-resolved
spectra, permitting precise measurements of the hard-X ray continuum from
accelerated electrons, of tens of strong nuclear lines and of the high-energy
emission from $\pi^0$ decay and $\pi^{\pm}$-decay leptons. This, for the
first time, would allow detailed studies of the evolution of the energy spectra
and composition of the accelerated particles which are the key properties for
the understanding of the acceleration mechanism and the transport of energetic
particles in solar flares.
In particular, it will be possible to determine the accelerated $^3$He/$^4$He
ratio and the heavy-ion content of the interacting particles and compare them
with observations of solar energetic particles in interplanetary space, where
large overabundances of $^3$He ($\approx$ 100 - 1000) and of low-FIP elements
($\approx$ 10-30) are found after impulsive-type flares. Additional information
would come from polarization measurements of the bremsstrahlung continuum and
of some nuclear lines and the detection of delayed X- and  $\gamma$-ray line
emission from solar active regions that are following the production of
relatively long-lived radio-isotopes during strong flares
(Tatischeff \etal\ 2006).

\section{Mission profile}

For the GRIPS mission, 
a mass of 3.5\,t needs to be delivered to a circular equatorial low-Earth 
orbit (LEO), with an altitude of about 500\,km.
The Soyuz Fregat 2B is capable of launching this payload
(capacity of 5.3\,t) and with its fairing of diameter 3.8\,m and
height of 7.7\,m
readily accommodates GRIPS with the science instruments, the
Gamma-Ray (GRM) and X-Ray Monitors (XRM).

GRIPS will generally be operated in a continuous zenith pointed
scanning mode. The field of view (diameter 160$\degs$) will cover
most of the sky over the course of one orbit. A similar strategy
is planned for the all-sky survey of the high-energy telescope LAT
on the forthcoming GLAST mission. Pointing the XRT to a selected
source will not disturb the scanning coverage of the GRM, since
the detector response is nearly invariant under rotation around
the axis.

\section{Payload}

\subsection{Overview of instruments}

GRIPS will carry two major telescopes: the Gamma-Ray Monitor (GRM)
and the X-Ray Monitor (XRM). The GRM is a combined Compton
scattering and Pair creation telescope for the energy range
0.2--50 MeV. It will thus follow the successful concepts of
imaging high-energy photons used in COMPTEL (0.7--30 MeV) and
EGRET ($>$30 MeV) but combines them into one instrument. New
detector technology and a design that is highly focused on the
rejection of instrumental background will provide greatly improved
capabilities. Over an extended energy range the sensitivity will
be improved by at least an order of magnitude with respect to previous 
missions (Fig. \ref{sens}) and the large field of view, better angular and
spectroscopic resolution of GRM allows the scientific goals
outlined in this project to be addressed. The XRM is based on the
mature concept and components of the eROSITA X-ray telescope,
which is scheduled for a space mission on the Russian platform
Spektrum-XG  (Predehl \etal\ 2006, Pavlinsky \etal\ 2006).


\end{document}